\documentclass[twocolumn,prl,aps,showpacs,10pt,superscriptaddress]{revtex4-1}
\usepackage{amsfonts}
\usepackage{amsmath}
\usepackage{mathtools}
\usepackage{rotating}
\usepackage{amssymb}
\usepackage{graphicx}
\usepackage{bm}
\usepackage{xcolor}
\usepackage{tabularx}
\newcommand{\mbf}[1]{\mathbf{#1}}

\renewcommand{\t}[1]{\textrm{#1}}

\newcommand{\nn}[0]{\nonumber\\}

\begin{document}

\title{Insensitivity of spin dynamics to the orbital angular momentum transferred from twisted light 
to extended semiconductors}

\author{M.~Cygorek}
\affiliation{Theoretische Physik III, Universit{\"a}t Bayreuth, 95440 Bayreuth, Germany}

\author{P.~I.~Tamborenea}
\affiliation{Departamento de F\'isica and IFIBA, FCEN, Universidad de Buenos Aires, Ciudad
Universitaria, Pabell\'on I, 1428 Ciudad de Buenos Aires, Argentina}
\affiliation{Theoretische Physik III, Universit{\"a}t Bayreuth, 95440 Bayreuth, Germany}

\author{V.~M.~Axt}
\affiliation{Theoretische Physik III, Universit{\"a}t Bayreuth, 95440 Bayreuth, Germany}

\begin{abstract}
We study the spin dynamics of carriers due to the Rashba interaction in semiconductor quantum 
disks and wells after excitation with light with orbital angular momentum. 
We find that although twisted light transfers orbital angular momentum to the excited carriers 
and the Rashba interaction conserves their total angular momentum, the resulting electronic 
spin dynamics is essentially the same for excitation with light with orbital angular 
momentum $l=+|l|$ and $l=-|l|$. 
The differences between cases with different values of $|l|$ are due to 
the excitation of states with slightly different energies and not to the 
different angular momenta per se, and vanish for samples with large radii 
where a $k$-space quasi-continuum limit can be established. 
These findings apply not only to the Rashba interaction but also to all other envelope-function 
approximation spin-orbit Hamiltonians like the Dresselhaus coupling. 
\end{abstract}

\maketitle


Light with orbital angular momentum (OAM), referred to as twisted light, is a relatively new
field of research which has become increasingly popular 
\cite{twistedlightXiang14,twistedlightDeNinno14,twistedlightCorkum14,
twistedlightPadgett14,
twistedlightCouprie13,twistedlightSpanner12,twistedlightWillner12,
twistedlightArie12,
twistedlightCapasso12,twistedlightSerbo11,twistedlightBowman11,
twistedlightUnguris11,
twistedlightSchattschneider10,twistedlightTonomura10,
twistedlightTorner07,twistedlightPadgett02,twistedlightRubinsztein95} 
since Allen \textit{et al.}\ showed how twisted light beams can
be easily generated from conventional laser beams \cite{twliAllen}.
Recently, the theoretical foundation of the optical excitation of solids and nanostructures
with twisted light has been established
\cite{twliEPL09,twliPRB09,twliOE09,twliPRB10,twliOE11,twliBerakdar12,twliCM13,
twliGraphene13,twliKuhn14}, 
and experimental studies with twisted light on semiconductors
have been carried out \cite{twliOEexp09,twliGay}.

One motivation for such studies is the prospect of using the large amounts of 
angular momentum that twisted light can carry in order to control the spin dynamics 
of electrons, thus adding a flexible tool to the active field of spin control
\cite{optorient,spincontrolFlensberg06,spincontrolPrettl01,
spincontrolPeeters12,Zutic,spincontrolBeschoten10,spincontrolBakarov14,
spincontrolBeschoten12,spincontrolGershoni11,spincontrolBayer09}.
In this context two different mechanisms need to be distinguished.
First, angular momentum as well as energy selection rules can lead to 
selective optical excitation of carriers with a preferred spin direction.
This mechanism enables fast spin-selective preparation of states during
the photoexcitation process and has recently been studied for strongly confined systems 
such as quantum dots \cite{twliKuhn14} and quantum rings \cite{twliOE11}.
Secondly, the spin-orbit interaction---like the Rashba \cite{Rashba} and  
Dresselhaus \cite{Dresselhaus} couplings in semiconductor structures---is expected 
to couple the OAM of carriers transferred from the twisted light \cite{twliEPL09,twliPRB10} 
to their spin degree of freedom.
This would provide a slower carrier spin control which would be dynamical 
and would remain active after the twisted light pulse.

In this Letter, we study the spin dynamics of carriers in semiconductor quantum disks and wells 
excited with twisted light taking into account the Rashba spin-orbit interaction. 
Our central finding is that, rather unexpectedly, the spin dynamics of the photo-excited 
electrons differs only slightly after excitation with light with and without OAM 
in the limit of large quantum disks, becoming insensitive to the OAM content 
of the twisted light beam for extended quantum wells.
This result is consistent with the outcome of recent experiments
which did not show traces of the OAM transferred from twisted light to bulk GaAs
in spin-resolved photoemission measurements \cite{twliGay}.

Analytically, we find that the Rashba interaction, while conserving the
total angular momentum of the electrons, has matrix elements
which are independent of the OAM quantum number in the (k-space)
quasi-continuum limit.
As a consequence, the induced spin dynamics is almost identical, in particular, 
for twisted light with components of the OAM in the growth direction $l=+|l|$ and $l=-|l|$.
This finding can be generalized to all possible effective spin-orbit 
interactions stemming from a lattice-periodic potential 
in the envelope-function approximation, e.~g.\ the Dresselhaus coupling. 
From this we conclude that the dynamical spin control mechanism 
as analyzed here can only be effective for small quantum disks 
and other strongly confined systems.

\begin{figure*}[t!]
\includegraphics{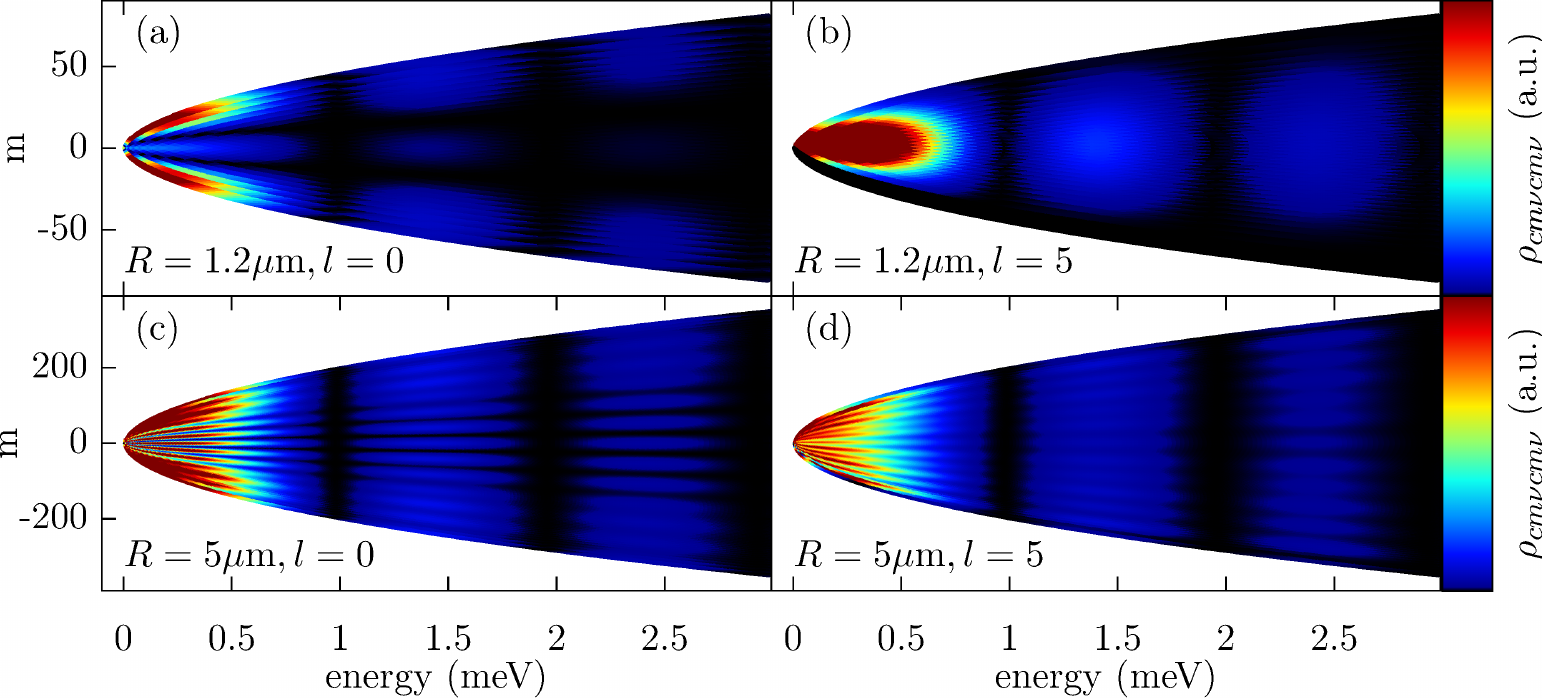}
\caption{Diagonal elements of the density matrix $\rho_{cm\nu cm\nu}$ 
after an excitation of a cylindrical quantum disk with radius $R$ 
with a box-shaped pulse of length $t=5$ ps and orbital angular momentum $l$ of the light.
(a) and (b) show the occupations of a $R=1.2\;\mu\t{m}$ quantum disk
with $l=0$ and $l=5$, respectively. 
(c) and (d) display the occupations for a larger $R=5\;\mu\t{m}$ disk with $l=0$ and $l=5$.
For a better comparison, the matrix elements are plotted against the energies $\epsilon_{c{m\mu}}$ 
instead of the indices $\mu$ and the absolute values of the occupations are rescaled.}
\label{fig:occup}
\end{figure*}

The discussion of the optical excitation of electrons with twisted light
is especially clear when a basis of cylindrical states is chosen \cite{twliPRB10}.
The wave functions of these basis states are expressed in cylindrical 
coordinates $\{r,\phi,z\}$ as
\begin{align}
  \psi_{bm\nu}(r,\phi,z) = \mathcal{N}_{m\nu} J_m(k_{m\nu}r) e^{im\phi} \Phi_b(z),
\label{eq:cylstat}
\end{align}
where $J_m$ is the $m$-th Bessel function, $\Phi_b(z)$ is the $z$-envelope of the $b$ subband 
(the band index includes the spin quantum number) and 
$\mathcal{N}_{m\nu}=[\sqrt\pi RJ_{m+1}(k_{m\nu}R)]^{-1}$
is the normalization factor. 
For a circular quantum disk with radius $R$, width $L$ and growth direction $z$,
the boundary conditions $\psi_{bm\nu}(R,\phi,z)=0$ are satisfied, if
$k_{m\nu}=u_{m,\nu}/R$ where $u_{m,\nu}$ is the $\nu$-th zero of 
the $m$-th Bessel function.
Note that $\psi_{bm\nu}$ is an eigenstate of the $z$-component of the envelope OAM operator 
with eigenvalue $\hbar m$ and $k_{m\nu}$ determines the kinetic energy of the state $\psi_{bm\nu}$, 
since in a parabolic band $b$ with effective mass $m_b^*$ the energy of the state 
is given by $\epsilon_{b{m\nu}}=\hbar^2k_{m\nu}^2 / (2m_b^*)$.
Note also how the precise location of the energy eigenvalues is given by the zeros of the Bessel 
functions; this detailed information will be ``smeared out'' in the limit $R \rightarrow \infty$ 
as the allowed values of $k$ become a quasi-continuum.
For the sake of simplicity, we restrict our discussion to a case where only 
spin degenerate conduction ($b=c$) and heavy-hole ($b=v$) bands are considered.

The matrix elements of the twisted-light--matter interaction Hamiltonian $H_I$ 
(in the dipole approximation for only the $z$-component)
in the cylindrical basis states was derived in Ref.~\onlinecite{twliPRB10}:
\begin{subequations}
\begin{align}
& \langle c m'\nu'|H_I| v m\nu\rangle = 
    \xi_{cv\nu'\nu m'} e^{-i\omega t} \delta_{m,m'-l}, \\
&    \xi_{cv\nu'\nu m'} = \;
-\frac {e}{m_e} A_0 \, \boldsymbol{\epsilon}_{\sigma} \cdot \mbf p_{cv}
    \langle \Phi_c|\delta_{k_z,q_z} |\Phi_v\rangle
    \mathcal{N}_{m'\nu'}\mathcal{N}_{m\nu} \times \nn
& \hspace*{1.2cm} \times {\textstyle\int\limits_0^R} dr \;r 
                         J_l(q_\|r) J_{m'}(k_{m'\nu'}r) J_{m'-l}(k_{m\nu}r),
\end{align}
\label{eq:matelems}
\end{subequations}
where $e$ and $m_e$ are the electron charge and (bare) mass, 
$A_0$ is the field strength,
$\omega$ is the light frequency, 
$q_\|$ the in-plane and $q_z$ the growth-direction part of the light wave vector, 
$k_z$ is the electron wave vector in the growth direction,
$l$ the OAM of the light, 
$\boldsymbol{\epsilon}_{\sigma}$ is the light polarization vector and 
$\mbf p_{cv}$ is the dipole matrix element between heavy-hole and 
conduction band states. 
Note that $\mbf p_{cv}$ contains spin selection rules. 
Let us consider the case where due to excitation with circularly polarized light only 
spin-up electrons are excited.

In Ref.~\onlinecite{twliPRB10} equations of motion were presented for the density matrix 
under the influence of twisted light switched on with constant amplitude at $t=0$.
In the low-excitation limit, i.e.\ initially empty conduction and filled valence bands excited 
with a moderate light field so that the occupations can be well approximated
by an expansion up to second order in the field strength, we find from
Eqs.~(16) and (17) of Ref.~\onlinecite{twliPRB10}:
\begin{widetext}
\begin{align}
&\rho_{c m\nu c' m'\nu'}(t)=
    \sum_{m_1\nu_1v} \delta_{-\frac 32,v} \xi_{cv\nu\nu_1m} \xi_{c'v\nu'\nu_1m'}^*
    \bigg[\frac{1-e^{- i (\epsilon_{cm\nu} - \epsilon_{c'm'\nu'}) t/\hbar }}
               {\epsilon_{cm\nu}-\epsilon_{c'm'\nu'}}
          \bigg(
                \frac{1}{\epsilon_{c'm'\nu'}-\epsilon_{vm_1\nu_1}-\hbar\omega}
                -\frac{1}{\epsilon_{cm\nu}-\epsilon_{vm_1\nu_1}-\hbar\omega}
          \bigg)+
          \nn&+
          \frac{1}
               {(\epsilon_{cm\nu}-\epsilon_{vm_1\nu_1}-\hbar\omega)
                (\epsilon_{c'm'\nu'}-\epsilon_{vm_1\nu_1}-\hbar\omega)}
          \left(e^{i (\epsilon_{c'm'\nu'}-\epsilon_{vm_1\nu_1} - \hbar\omega) t / \hbar}
                +e^{-i (\epsilon_{cm\nu}-\epsilon_{vm_1\nu_1} - \hbar\omega) t / \hbar}-2
          \right)
    \bigg]
    \delta_{mm'} \delta_{c\frac12} \delta_{c'\frac 12}.
\label{eq:exc_dens_mat}
\end{align}
\end{widetext}

Thus, the optical excitation yields only diagonal elements of 
$\rho_{ckmc'k'm'}$ with respect to $m$, i.e.\ only states with a defined
envelope OAM are excited. 
Also, we find from Eqs.~(\ref{eq:matelems}) that for every electron with OAM $m$ 
a hole with OAM $m-l$ is excited.
From this we can conclude that the total envelope OAM $l^{tot}$ 
induced in valence and conduction band together is $l^{tot}=\hbar l N_e$ 
where $N_e$ is the number of excited electrons or, equivalently, holes.
For very short pulse times $t$, the diagonal elements of the density
matrix from Eq.~(\ref{eq:exc_dens_mat}) are given by 
\begin{align}
  \rho_{\frac 12 m\nu\frac 12 m\nu}\approx \sum_{\nu_1}
  |\xi_{\frac 12(-\frac32)\nu\nu_1m}|^2 \frac{t^2}{\hbar^2}.
\label{eq:veryshorttime}
\end{align}
Note that in Eq.~(\ref{eq:veryshorttime}) no information about the band 
structure is contained. In particular, the formula for the valence band
occupations is the same as for the conduction band. 
Thus, in this limit, the total angular momentum is distributed symmetrically to heavy-hole 
and conduction band, i.e.\ the total envelope OAM in the conduction
band is $l_{c}^{tot}=\hbar \frac{l}2 N_e$.
However, for longer pulse durations, the energy selection becomes important
leading to an in general different value of $l_{c}^{tot}$.

Figure \ref{fig:occup} shows the diagonal elements of the density matrix 
$\rho_{cm\nu cm\nu}$ after an excitation with pulse duration $t=5$ ps 
with circularly polarized light with OAM  $l$. 
The central frequency of the light pulse was chosen to be resonant with the band gap.
The effective masses were $m^*_c= 0.067 \, m_e$ and $m^*_{hh}=0.45 \, m_e$ 
for conduction and heavy-hole band, respectively.
The oscillatory structure of the occupations along the energy axis can be 
attributed to the finite pulse duration via the energy-time
uncertainty relation. 
Along the $m$ axis, there are also oscillations in the occupations. 
Since their frequency depends strongly on the radius $R$ of the sample 
and they get smeared out for large values of $R$, 
we attribute these oscillations to finite size effects.
Note that in Fig.~\ref{fig:occup}(b), where the occupation for light with
$l=5$ is plotted, the states with the 5 lowest values of $m$ for every energy 
shell are empty (seen more clearly at low energies), since there are no valence band states 
which satisfy the condition $m'=m-l$ of the matrix element in Eq.~(\ref{eq:matelems}a).
Figures \ref{fig:occup}(c) and (d) show that for the larger $R=5\;\mu$m quantum disk, 
the difference between the occupations after $l=0$ and $l=5$ excitations diminishes visibly.

\begin{figure}[t]
\includegraphics{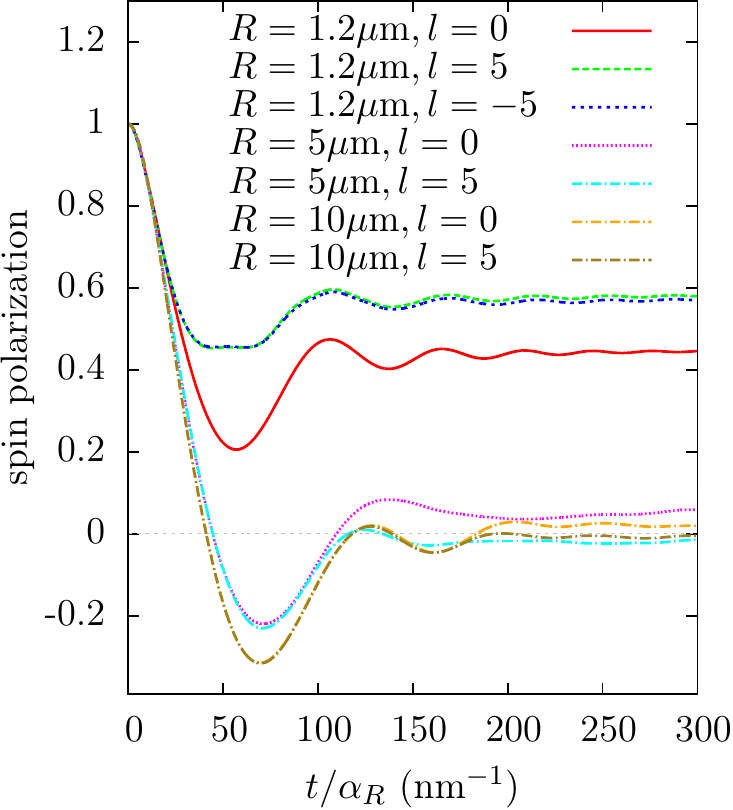}
  \caption{ Spin dynamics after excitation with twisted light with orbital angular momentum
            $l=\{-5, 0, 5\}$ and quantum-disk radius $R= 1.2 \, \mu \text{m}$
            and with $l=\{0, 5\}$ and $R= \{5 \, \mu \text{m}, 10 \, \mu \text{m} \}$.
          }
\label{fig:dyn}
\end{figure}

Now, we focus on the spin dynamics after the optical excitation. 
We study the effects of spin-orbit coupling mechanisms, considering for concreteness 
the Rashba Hamiltonian \cite{Rashba}, $H_R$, which is usually the dominant mechanism 
in quasi-two-dimensional systems.
In order to better work with the cylindrical states given in Eq.~(\ref{eq:cylstat}), 
we switch from the usual cartesian-coordinate expression of $H_R$ to its expression 
in polar coordinates: 
\begin{subequations}
\begin{align}
H_R=& \hbar\alpha_R \big(k_y\sigma_x-k_x\sigma_y\big)=
      \hbar\alpha_R \big(s^+\partial^- -s^-\partial^+\big),\\
      \partial^\pm:=& \frac\partial{\partial x}\pm i \frac\partial{\partial y}=
      e^{\pm i\phi}\bigg(\frac\partial{\partial r}\pm\frac ir 
      \frac\partial{\partial \phi}\bigg),
\end{align}
\label{eq:RashbaForm}
\end{subequations}
where $\alpha_R$ is the Rashba coefficient, $\sigma_i$ and $s^\pm$ are the 
Pauli matrices and spin raising and lowering operators, respectively.
With the relation 
\begin{align}
\label{eq:generalIntegral}
  &{\textstyle\int\limits_0^R}dr\;r J_{m}( pr)J_m(qr)=\nn&=
  R\frac{pJ_m(qR)J'_m(pR)-qJ_m(pR)J'_m(qR)}{q^2-p^2},
\end{align}
it is straightforward to calculate the matrix elements of $H_R$ with respect
to the cylindrical states:
\begin{subequations}
\begin{align}
  \langle c' m' \nu'|\partial^\pm|cm\nu\rangle &= \delta_{m',m\pm 1}
  \frac 2R\frac{k_{m\nu}k_{m'\nu'}}{k_{m\nu}^2-k_{m'\nu'}^2}, \\
  \langle c'm'\nu'|H_R|cm\nu\rangle &=\frac{\hbar\alpha_R k_{m\nu}k_{m'\nu'}
  }{R(k_{m\nu}^2-k_{m'\nu'}^2)}\times\nn&\times
  (s^+_{c'c}\delta_{m'\!,m-1}\!-\!s^-_{c'c}\delta_{m'\!,m+1}).
\end{align}
\label{eq:RashbaCyl}
\end{subequations}
It can be seen from the form of the Rashba Hamiltonian in cylindrical coordinates 
that an electron with spin-up (down) and envelope OAM $m$ flips to a state 
with spin-down (up) and OAM $m+1$ ($m-1$). 
In this sense, $\partial^\pm$ can be regarded as raising and lowering operators in $m$.
If $H_R$ is applied a second time, the electronic state is transferred back 
to the initial spin and OAM state, while a change in $\nu$ is possible. 
Note that the sum $J=m+s$ of the envelope OAM $m$ and the spin $s$ is conserved 
by the Rashba Hamiltonian.

Having derived the matrix elements of $H_R$ in cylindrical coordinates,
it is straightforward to calculate numerically the time evolution of the density matrix, 
where the initial conditions correspond to the final occupations generated by 
optical excitation with light with OAM $l$, illustrated in Fig.\ \ref{fig:occup}.
The resulting dynamics for the total conduction-band spin is shown in Fig.~\ref{fig:dyn}. 
We show results for three different values of the disk radius,
$R= \{ 1.2, 5, 10 \} \, \mu \text{m}$.
As in the case of optical excitation with light with zero OAM, the Rashba interaction 
leads to a dephasing of the initial electron spins. 
Since for small disks only a finite number of states contributes noticeably to the dynamics, 
oscillations are found which do not cancel completely so that for long times the total spin 
reaches a non-zero value.
Note that the curves for excitation with $l=5$ and $l=-5$ coincide (shown for 
$R= 1.2 \, \mu \text{m}$). 
This unexpected result shows clearly the insensitivity of the spin
dynamics, in the presence of the Rashba spin-orbit interaction, to the content 
of OAM transferred from twisted light to the electron gas.
For the same quantum disk, an excitation with $l=0$ produces spin dynamics slightly different 
from the $l=\pm 5$ excitation, but this difference decreases for larger radii. 
This tendency can be seen by comparing the excitation with $l=5$ and $l=0$ for the three
different values of $R$ used in Fig.~\ref{fig:dyn}.

To understand this finding, it is useful to analyze the case of an infinitely 
extended quantum well, obtained letting $R \rightarrow \infty$. 
In this limit, the discrete $k_{m\nu}$ become continuous and the eigenstates can be written as
\begin{align}
\psi_{b k m}(r,\phi,z):=
\sqrt{\frac{k}{2\pi}} J_m(k r) e^{im\phi} \Phi_b(z),
\label{eq:infstat}
\end{align}
with the orthogonality relation $\langle b km|b'k'm'\rangle=\delta_{bb'}\delta_{mm'}\delta(k-k')$.
Using that Bessel functions satisfy
\begin{align}
&{\textstyle\int\limits_0^\infty}dr\;r J_{m}( kr)J_m(k'r)=
\frac 1k \delta(k-k'),
\end{align}
the corresponding matrix elements become
\begin{subequations}
\begin{align}
   \langle c'k'm'|\partial^\pm|ckm\rangle &= \mp k\delta_{m',m\pm 1} \delta(k-k')\\
   \langle c'k'm'|H_R|ckm\rangle &=\hbar\alpha_R k\;\delta(k-k')\times\nn&\times
   (s^+_{c'c}\delta_{m'\!,m-1}\!+\!s^-_{c'c}\delta_{m'\!,m+1}).
\end{align}
\label{eq:InfRashbaCyl}
\end{subequations}
Thus, in the quasi-continuum limit, the prefactor of the Rashba-interaction
depends only on the energy of the state via $k$ but not on $m$.
The spin dynamics is therefore a precession of the electron spin with a $k$-dependent 
frequency which, for a given $k$, is the same for all different values of $m$.
Since the effect of the excitation with twisted light was mainly that
states with different $m$ are excited, it is now easy to see why, for extended systems, 
the spin dynamics due to the Rashba Hamiltonian is almost the same for excitations 
with light with and without OAM.

It is noteworthy that this statement is also true for all effective Hamiltonians 
with a microscopic origin in the lattice-periodic crystal potential such as the 
Dresselhaus \cite{Dresselhaus} spin-orbit coupling when treated in the envelope-function 
approximation:
For a lattice-periodic potential, the solutions of the corresponding Schr\"odinger 
equation are given by the Bloch theorem as 
$\psi(\mbf r)\propto e^{i\mbf k\mbf r}u_{n\mbf k}(\mbf r)$ 
with lattice-periodic Bloch function 
$u_{n\mbf k}(\mbf r)$, band index $n$ and wave-vector $\mbf k$.
The envelope-function approximation consists of integrating over the plane-wave 
part of the wave function yielding an effective Hamiltonian \cite{Bastard_kp} 
$H_{\text{eff}}$ for the $u_{n\mbf k}$ which is diagonal in $\mbf k$
and the matrix elements can be written as a power series in $\mbf k$. 
The resulting effective Hamiltonian can be rewritten by decomposing $k_x$ and $k_y$ 
in terms of $\partial^+$ and $\partial^-$, as done in Eqs.~(\ref{eq:RashbaForm}).
Thus, the dependence of the matrix elements of $H_{\text{eff}}$ in cylindrical
states on $m$ is of the same character as for the Rashba Hamiltonian and
vanishes in the quasi-continuum limit.
For systems with finite size, however, a weak dependence on $m$ can be found due 
to the $m$-dependence of the possible $k$-values in the prefactor of the Rashba 
Hamiltonian in Eq.~(\ref{eq:RashbaCyl}b). 
This means that, e.g.\ for small quantum disks, where the energy separation between 
the discrete cylindrical states becomes important, the OAM of the exciting light 
can influence the spin dynamics significantly.

In conclusion, we have shown that, although the orbital momentum 
of light can be transferred into the envelope orbital angular momentum of electrons, 
the usual solid state spin-orbit interactions, such as Rashba and Dresselhaus
interactions, do not couple the envelope orbital momentum of the carriers
to the spin degree of freedom in such a way that a significant difference
in the spin dynamics after excitation with light with and without orbital 
momentum is found in large extended systems. 
This finding can explain that in recent experiments \cite{twliGay} no influence of the 
light orbital angular momentum on the spin polarization was found.
However, for cylindrical quantum disks with small radii,
the discreteness of the states plays an important role so that
the spin dynamics indeed depends on the orbital momentum of the light.
Nevertheless, also for small systems, the spin dynamics
after excitation with orbital momentum $l=|l|$ and $l=-|l|$ 
are very similar, in contrast to optical excitation with opposite circular
polarization, where the spin dynamics acquires a different sign.

\begin{acknowledgments}
We gratefully acknowledge the financial support of the 
Deutsche Forschungsgemeinschaft through grant No.\ AX17/9-1.
Financial support was also received from the Universidad de
Buenos Aires, project UBACyT 2011-2014 No. 20020100100741, 
and from CONICET, project PIP 11220110100091.
\end{acknowledgments}

\bibliography{april2015}

\end{document}